    \documentclass[prd,aps,preprint,tightenlines,showpacs,nofootinbib,superscriptaddress]{revtex4-1}
\usepackage{mathrsfs}
\usepackage{amsfonts}
\usepackage{amsmath}
\usepackage{amssymb}
\usepackage{array}
\usepackage{verbatim}
\usepackage{bm}
\usepackage{epsfig}
\usepackage{graphicx,color}
\usepackage{relsize}
\usepackage{lineno}
\usepackage{float}
\usepackage{multirow}
\RequirePackage{xspace}

\def\lsim{\raise0.3ex\hbox{$<$\kern-0.75em\raise-1.1ex\hbox{$\sim$}}}

\def\gsim{\raise0.3ex\hbox{$>$\kern-0.75em\raise-1.1ex\hbox{$\sim$}}}

\newcommand{\be}{\begin{equation}}

\newcommand{\ee}{\end{equation}}

\def\beq{\begin{equation}}

\def\eeq{\end{equation}}

\def\beqa{\begin{eqnarray}}

\def\eeqa{\end{eqnarray}}

\newcommand{\ba}{\begin{eqnarray}}

\newcommand{\ea}{\end{eqnarray}}

\def\gappeq{\mathrel{\rlap {\raise.5ex\hbox{$>$}}

{\lower.5ex\hbox{$\sim$}}}}

\def\lappeq{\mathrel{\rlap{\raise.5ex\hbox{$<$}}

{\lower.5ex\hbox{$\sim$}}}}

\def\Toprel#1\over#2{\mathrel{\mathop{#2}\limits^{#1}}}

\begin{document}

\title{Estimating the BFKL effects on the vector meson + jet production in photon - induced interactions at the LHC}

\vspace{3cm}

\author{Jo\~ao V.  {\sc Bulh\~oes}}
\email{joaovitor1729@gmail.com}
\affiliation{Institute of Physics and Mathematics, Federal University of Pelotas, \\
  Postal Code 354,  96010-900, Pelotas, RS, Brazil}

\author{Victor P. {\sc Gon\c{c}alves}}
\email{barros@ufpel.edu.br}
\affiliation{Institute of Physics and Mathematics, Federal University of Pelotas, \\
  Postal Code 354,  96010-900, Pelotas, RS, Brazil}

\author{Werner K. {\sc Sauter}}
\email{werner.sauter@ufpel.edu.br}
\affiliation{Institute of Physics and Mathematics, Federal University of Pelotas, \\
  Postal Code 354,  96010-900, Pelotas, RS, Brazil}

\vspace{3cm}

\begin{abstract}
The associated vector meson + jet production in photon - induced interactions at the LHC is investigated and predictions for the  cross - sections are derived considering the NLO corrections to the BFKL kernel.  We explore the possibility that in a near future the FOCAL detector  could be used to measure the  jet at forward rapidities, with the associated meson being measured by a central or forward detector, and estimate the rapidity distributions for the $\rho$ + jet and $J/\Psi$ + jet photoproduction in $pp$, $Pbp$ and $PbPb$ collisions. We demonstrate that the associated cross-sections are non - negligible and that a future experimental analysis  can be useful to constrain the description of the BFKL kernel and improve our understanding of the QCD dynamics.

\end{abstract}



\maketitle

\vspace{1cm}

\section{Introduction}

The description of the high energy limit of  strong interactions theory is one of the challenges of the Standard Model \cite{Lipatov:1996ts}. Theoretically, at ultra-high energies, unitarity corrections are expected to determine the behavior of the cross - sections \cite{hdqcd}. On the other hand, at large hard scales, the standard DGLAP framework of the QCD evolution \cite{dglap} is expected to remain valid. In a limited kinematical window between these limits, the observables are expected to be described by the BFKL approach \cite{Lipatov:1976zz,Kuraev:1976ge,Kuraev:1977fs,Balitsky:1978ic,Lipatov:1985uk}, which was proposed in the late 1970s by Lipatov and collaborators to describe the Regge limit of hadronic reactions. Such an expectation has motivated the experimental analysis of  processes characterized by the production of two  particles separated by a large rapidity gap, which are expected to be sensitive to the high - energy dynamics \cite{Mueller:1986ey,Mueller:1992pe,DelDuca:1993mn,Stirling:1994he,Andersen:2001kta,Motyka:2001zh,Enberg:2001ev}, and the calculation of the next - to - leading order (NLO) corrections to the BFKL approach \cite{Fadin:1998py,Ciafaloni:1998gs,Salam:1998tj,Ciafaloni:1999yw,Brodsky:1998kn}. In recent years, a large number of  phenomenological studies related to the probing of the BFKL dynamics in hadronic collisions at the LHC have been performed \cite{Chevallier:2009cu,Colferai:2010wu,Kepka:2010hu,Ivanov:2012iv,Ducloue:2013hia,Ducloue:2013bva,Hentschinski:2014bra,Hentschinski:2014esa,
Caporale:2014gpa,Celiberto:2015yba,Celiberto:2016hae,Celiberto:2017ptm,Boussarie:2017oae,Bolognino:2018oth,Golec-Biernat:2018kem,Celiberto:2022dyf,Baldenegro:2022kvj,Celiberto:2020wpk,Celiberto:2022gji,Egorov:2023duz,Colferai:2023hje}. Additionally, the study of photon - induced interactions at the LHC became a reality \cite{upc}, allowing us to investigate the impact of the QCD dynamics on the photon - hadron cross - sections at  center-of-mass energies that are at least one order of magnitude larger than those reached in  $ep$ colliders. 

\begin{figure}[t]
	\centering
\includegraphics[width=0.5\textwidth]{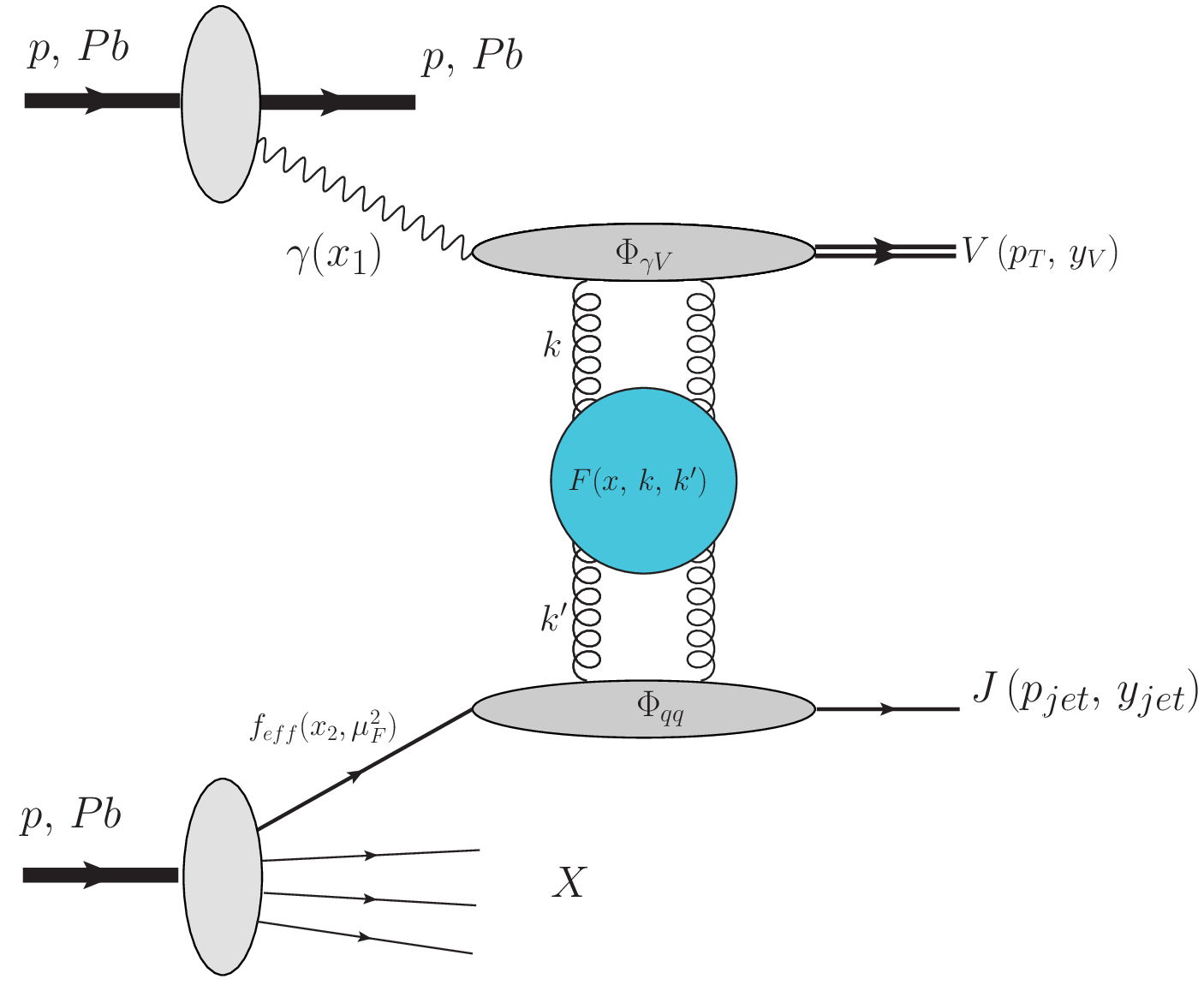} 
\caption{Associated vector meson + jet photoproduction in hadronic collisions. }
\label{fig:diagram}
\end{figure}

One of the promising processes to probing the BFKL dynamics is the associated vector meson + jet photoproduction in hadronic collisions, represented in Fig. 
\ref{fig:diagram}, which is expected to dominate the photoproduction of vector mesons at large momentum transfer. Such a process was previously investigated in Refs. \cite{Frankfurt:2006tp,Frankfurt:2008er,Goncalves:2009gs,Goncalves:2010zb,Cisek:2016kvr,Goncalves:2017sek}   considering different approximations for the description of the BFKL dynamics at leading order (LO). Differently from these studies, we will estimate the cross - sections considering the NLO corrections to the BFKL kernel, represented by the function $F(x,k,k^{\prime})$ in Fig. \ref{fig:diagram}. Moreover, we will not integrate over the jet kinematics, but instead to assume that it is measured. In our analysis, we will focus on the case that the jet is produced in the rapidity range that will be covered by the FOCAL detector ($3.2 < y < 5.8$) \cite{ALICE:2024jtt,Jonas:2024ftk} and  that the vector meson is measured by the ALICE central detectors or also by the FOCAL\footnote{The potentiality of the FOCAL detector to probe photon - induced interactions has been discussed in Ref. \cite{Bylinkin:2022wkm}, which we refer to the interested reader.}.  In this process, the final state will be characterized by two rapidity gaps, being one between the vector meson and the hadron that
emits the photon and remains intact, and the other gap between the vector meson and the jet. Moreover, we will assume that the $t$-channel color singlet object  exchanged between the vector meson and the jet carries large momentum transfer, which implies that the hadron target will dissociate. We will consider $pp$, $pPb$ and $PbPb$ collisions and in order to demonstrate the sensitivity to the BFKL dynamics, the NLO predictions will be compared with the LO and Born level results. As we will demonstrate below, a future experimental analysis of this process can be useful to improve our understanding of the QCD dynamics at high energies.

\section{Formalism}

Initially, let's present a brief review of the formalism needed to describe the associated vector meson + jet photoproduction in hadronic collisions at high energies (For a review see, e.g., Ref. \cite{Enberg:2004rz}). In our analysis, we will assume that the intact hadron in the final state can be tagged, which implies the identification of the photon source. Considering a generic hadronic collision $h_1 h_2 \rightarrow h_1 \otimes V \otimes J X$, with $\otimes$ representing a rapidity gap in the final state, $h_i = p, \, Pb$ and $V = \rho,\, J\Psi$, one has that the associated cross-section can be expressed as follows
\begin{eqnarray}
\frac{d\sigma_{h_1 h_2 \rightarrow h_1 \otimes V \otimes J X}}{dy_{V}d^2p_Tdy_{Jet}} =  x_1 \gamma^{h_1}(x_1) x_2f^{h_2}_{eff}(x_2, \mu^2_F)\, \frac{d\sigma(\gamma f \rightarrow V f)}{dt} \,\,\,,
\label{Eq:dif}
\end{eqnarray} 
where $|t| =  p_T^2$ is the squared transverse momentum carried by the color - singlet object exchanged between the vector meson and the jet, $x_1\gamma^{h_1}$ represents the elastic photon distribution associated with the hadron $h_1$ and $x_2f^{h_2}_{eff}$ is the effective parton distribution of the hadron $h_2$, defined by
\begin{eqnarray}
    x_2f^{h_2}_{eff}(x_2, \mu_F^2)=\frac{81}{16}x_2g^{h_2}(x_2,\mu_F^2)+\sum_{f}[x_2q^{h_2}_f(x_2,\mu_F^2)+x_2\Bar{q}^{h_2}_f(x_2,\mu_F^2)]\,\,,
\end{eqnarray}
with     $\mu^2_F=m_V^2+p_T^2$. Moreover, $\hat{s} = x_1 x_2 s$, with $\sqrt{s}$ being the center - of - mass energy of the $h_1 h_2$ collision, and the longitudinal momentum fractions given by
\begin{eqnarray}
    x_1=\sqrt{\frac{M^2_{V}+p_T^2}{s}}(e^{y_{V}}+e^{y_{Jet}}) \,\,\,\mbox{and} \,\,\, x_2=\sqrt{\frac{M^2_{V}+p_T^2}{s}}(e^{-y_{V}}+e^{-y_{Jet}}).
\end{eqnarray}
As in previous studies \cite{Forshaw:2001pf,Enberg:2003jw,Poludniowski:2003yk}, we will write the differential cross - section for the partonic subprocess as follows
\begin{eqnarray}
    \frac{d\sigma(\gamma f\rightarrow V f)}{dt}=\frac{16\pi}{81t^4}|\mathcal{F}(z, \tau)|^2 \,\,,
\end{eqnarray}
with the function $\mathcal{F}$ being expressed by
\begin{equation}
    \mathcal{F}(z,\tau)=\frac{t^2}{(2\pi)^2}\int d\nu\frac{\nu^2}{(\nu^2+1/4)^2}e^{\omega(\nu)z}I_{\nu}^{\gamma V}(\sqrt{|t|})I_{\nu}^{qq}(\sqrt{|t|})^* \,\,,
    \label{Eq:efe}
\end{equation}
where  $\tau={|t|}/{(M^2_{V}+Q^2_{\gamma})}$ and $z=({3\alpha_s}/{2\pi})\ln{\left({s/}{\Lambda^2}\right)}$, with $Q_{\gamma}^2$ being the photon virtuality and $\Lambda^2$ a characteristic scale related to $M_V^2$ and $|t|$. As in Refs. \cite{Goncalves:2009gs,Goncalves:2010zb,Goncalves:2017sek,Forshaw:2001pf} we will assume $\Lambda^2 = \beta M_V^2 +\delta |t|$, with the constants $\beta$ and $\delta$ being fixed using the HERA data \cite{ZEUS:2002vvv,H1:2003ksk}. It is important to emphasize that in photon - induced interactions at the LHC, the photon virtuality is very small and assume 
$Q_{\gamma}^2 = 0$ is a very good approximation \cite{upc}. 
{ The quantities  $I_{\nu}^{\gamma V}$  and  $I_{\nu}^{qq}$ are given in terms of the BFKL eigenfunctions and of the photon - vector meson ($\mathcal{I}_{\gamma V}$)  and quark - quark ($\mathcal{I}_{qq}$) impact factors as follows \cite{Enberg:2003jw,Poludniowski:2003yk}
\begin{eqnarray}
I_{\nu}^{ab}(Q_{\perp }) & = & \int \frac{d^{2}k_{\perp }}{(2\pi )^{2}}\, {\mathcal{I}}_{ab}(k_{\perp },Q_{\perp })\int d^{2}\rho _{1}d^{2}\rho _{2} \label{inugv}\\
 & \times  & \biggl [\left( \frac{(\rho _{1}-\rho _{2})^{2}}{\rho _{1}^{2}\rho _{2}^{2}}\right) ^{1/2+i\nu }-\left( \frac{1}{\rho _{1}^{2}}\right) ^{1/2+i\nu }-\left( \frac{1}{\rho _{2}^{2}}\right) ^{1/2+i\nu }\biggr ]e^{ik_{\perp }\cdot \rho _{1}+i(Q_{\perp }-k_{\perp })\cdot \rho _{2}}.\nonumber 
\end{eqnarray}
In the case of coupling to a colorless state, only the first term in the square bracket { remains} since \( {\mathcal{I}}_{ab}(k_{\perp },Q_{\perp }=k_{\perp })={\mathcal{I}}_{ab}(k_{\perp }=0,Q_{\perp })=0 \). 
The impact factor ${\mathcal{I}}_{\gamma V}$ describes, in the high energy limit, the couplings of the external particle pair to the color singlet gluonic ladder. Following previous studies \cite{Forshaw:2001pf,Enberg:2003jw,Poludniowski:2003yk}, we will assume that \cite{Ryskin:1992ui}:
\begin{eqnarray}
{\mathcal{I}}_{\gamma V}\;=\;
\frac{{\mathcal C} \alpha_s}{2}\, \biggl(\frac{1}{\bar{q}^{2}}-
\frac{1}{q_{\|}^{2}+k_{\perp}^{2}} \biggr), 
\label{impfmom}
\end{eqnarray}
where
\begin{eqnarray}
\bar{q}^{2}\;=\;q_{\|}^{2}+Q_{\perp}^{2}/4, \\
q_{\|}^{2}\;=\;(Q_{\gamma}^{2}+M_{V}^{2})/4,
\label{C}
\end{eqnarray}
and  the constant ${\mathcal C}$ {can} be expressed in terms of the vector meson leptonic decay width as follows
\begin{eqnarray}
\mathcal{C}^{2}\;=\;\frac{3\Gamma_{ee}^{V}M_{V}^{3}}{\alpha_{\mathrm{em}}}.
\end{eqnarray}
In the calculation of ${\mathcal{I}}_{\gamma V}$ , the factorization of the scattering process and the meson formation is assumed, as well as the validity of a non-relativistic approximation of the meson wave function. 
Using Eq. (\ref{impfmom}) into (\ref{inugv}), results that the function   $I_{\nu}^{\gamma V}(Q_{\perp })$ is given by \cite{Forshaw:2001pf,Enberg:2003jw,Poludniowski:2003yk}
\begin{eqnarray}
I_{\nu}^{\gamma V}(Q_{\perp }) & = & -{\mathcal{C}}\, \alpha_s \frac{16\pi}{Q_{\perp }^{3}}\frac{\Gamma (1/2-i\nu )}{\Gamma (1/2+i\nu )}\biggl (\frac{Q_{\perp }^{2}}{4}\biggr )^{i\nu }\int _{1/2-i\infty }^{1/2+i\infty }\frac{du}{2\pi i}\biggl (\frac{Q_{\perp }^2}{4 M_{V}^2}\biggr )^{1/2+u}\\
 &  & \times\frac{\Gamma ^{2}(1/2+u)\Gamma (1/2-u/2-i\nu/2)\Gamma (1/2-u/2+i\nu/2)}{\Gamma (1/2+u/2-i\nu /2)\Gamma (1/2+u/2+i\nu /2)}.\nonumber \label{IV} 
\end{eqnarray}
On the other hand, the quark impact factor is given by ${\mathcal{I}}_{q q} = \alpha_s$, which implies \cite{Bartels:1996fs}
 \begin{eqnarray}
I_{\nu}^{q q}(Q_{\perp }) & = & - \frac{4\pi \alpha_s}{Q_{\perp }} \left(\frac{Q_{\perp }^2}{4}\right)^{i\nu} \frac{\Gamma(\frac{1}{2}-i\nu)}{\Gamma(\frac{1}{2}+i\nu)}\,\,.
\label{iq}
\end{eqnarray}
}

{ The function $ \omega(\nu)$ in Eq. (\ref{Eq:efe}) is the BFKL characteristic function}, which determines the energy dependence of the cross - section, and is given by $ \omega(\nu)=\Bar{\alpha}_s\chi(\gamma)$, with  $\gamma=1/2+i\nu$ and $\Bar{\alpha}_s=(N_c\alpha_s)/\pi$. At leading order, the function $\chi(\gamma)$ is given by
\begin{equation}
    \chi^{LO}(\gamma)=2\psi(1)-\psi(\gamma)-\psi(1-\gamma)\,\,,
\end{equation}
with $\psi(x)$ being the digamma function. The leading order expressions for the impact factors and characteristic function were used in Refs. \cite{Goncalves:2009gs,Goncalves:2010zb,Goncalves:2017sek}   to estimate the vector meson photoproduction at large momentum transfer in hadronic collisions at the LHC.  In order to improve these previous estimates, here we will consider the NLO BFKL characteristic function, derived in Ref. \cite{Brodsky:1998kn} using the Brodsky-Lepage-Mackenzie (BLM) optimal scale setting  and the momentum space subtraction (MOM) scheme of renormalization \cite{Brodsky:1982gc,Brodsky:2002ka,Brodsky:2011ig}. In this approach, the BFKL characteristic function is given by
\begin{equation}
    \omega_{BLM}^{MOM}=\chi^{LO}(\gamma)\frac{\alpha_{MOM}(\hat{Q}^2)N_c}{\pi}\left[1+\hat{r}(\nu)\frac{\alpha_{MOM}(\hat{Q}^2)}{\pi}\right]
\end{equation}
where $\alpha_{MOM}$ is the coupling constant in the MOM scheme,
\begin{equation}
    \alpha_{MOM}=\alpha_s\left[1+\frac{\alpha_s}{\pi}T_{MOM}\right] \,\,\,\,\mbox{and} \,\,\,\, \alpha_s(\mu^2)=\frac{4\pi}{\beta_0\ln(\mu^2/\Lambda_{QCD}^2)} \,\,,
\end{equation}
with { $T_{MOM}$ expressed as follows \cite{Brodsky:1998kn}
\begin{eqnarray}
T_{MOM}& = & \frac{N_C}{8} \biggl[ \frac{17}{2} I + 
\xi \frac{3}{2} (I-1) + \xi^2 (1-\frac{1}{3}I) - 
\xi^3 \frac{1}{6} \biggr]  - \frac{ \beta_0}{2} \biggl[ 1 +\frac{2}{3} I \bigg] , 
\nonumber
\end{eqnarray}
where $I=-2 \int^{1}_{0}dx \ln(x)/[x^2-x+1]\simeq 2.3439$ and $\xi$ is a gauge parameter. In our calculations, we assume the Yennie gauge: $\xi = 3$.}
 The function $\hat{Q}$ is the BLM optimal scale and is given by
\begin{equation}
    \hat{Q}^2(\nu)=Q^2\exp\left[\frac{1}{2}\chi^{LO}(\gamma)-\frac{5}{3}+2\left(1+\frac{2}{3}I\right)\right].
\end{equation}
Finally, $\hat{r}$ is the NLO coefficient of the characteristic function, 
\begin{equation}
    \begin{split}
        \hat{r}(\nu)=&-\frac{\beta_0}{4}\left[\frac{\chi^{LO}(\nu)}{2}-\frac{5}{3}\right]-\frac{N_c}{4\chi^{LO}(\nu)}\left\{\frac{\pi^2\sinh(\pi\nu)}{2\nu \cosh^2(\pi\nu)}\left[3+\left(1+\frac{N_f}{N_c^3}\right)\frac{11+12\nu^2}{16(1+\nu^2)}\right]\right.\\&\left.-\chi''^{LO}(\nu)+\frac{\pi^2-4}{3}\chi^{LO}(\nu)-\frac{\pi^3}{\cosh(\pi\nu)}-6\zeta(3)+4\tilde{\phi}(\nu)\right\}+7.471-1.281\beta_0
    \end{split}
\end{equation}
with
\begin{equation}
    \tilde{\phi}(\nu)=2\int_0^1 dx\frac{\cos(\nu \ln (x))}{(1+x)\sqrt{x}}\left[\frac{\pi^2}{6}-\text{Li}_2(x)\right].
\end{equation}
In our analysis we will estimate the cross - sections using the LO impact factors, which implies that our calculation is not a full NLO one. In principle, the NLO corrections for the impact factors are expected to modify the normalization of the predictions, with the energy dependence being determined by the NLO BFKL characteristic functions. As the normalization of our predictions will be fixed using the HERA data, we believe that part of these NLO corrections for the impact factors will be included in our analysis. Surely, a full NLO calculation is an important next step, which we intend to perform in a forthcoming study.

\section{Results}

In what follows we will present our predictions for the associated vector meson + jet photoproduction in $pp$, $Pbp$ and $PbPb$ collisions at $\sqrt{s} = 14$ TeV, 8.16 TeV and 5.5 TeV, respectively. We will consider $\rho$ and $J/\Psi$ production, assuming that the elastic photon distribution associated with the proton and nucleus are given by \cite{upc}
\begin{equation}
    \gamma^p(x)=\frac{\alpha_{em}}{2\pi x}[1+(1-x)^2]\left[\ln{\Omega}-\frac{11}{6}+\frac{3}{\Omega}-\frac{3}{2\Omega^2}+\frac{1}{3\Omega^3}\right] \,\,,
\end{equation}
and
\begin{equation}
    \gamma^A(x)=\frac{Z^2 \alpha_{em}}{\pi} \frac{1}{x} [2 \xi K_0(\xi) K_1(\xi )-\xi^2(K_1^2(\xi )- K_0^2(\xi )) ]\,\,,
\end{equation}
with $\Omega=1+{0.71}/{Q^2_{min}}$, $Q^2_{min}={m_p^2x^2}/({1-x})$ and $\xi=x M_A b_{min}$, where $b_{min}$ is the sum of the radius of incident hadrons. For the  parton distributions functions (PDFs) of the proton, we will assume the MMHT2014LO parametrization \cite{Harland-Lang:2014zoa} and the nuclear PDFs will be calculated taking into account of the nuclear effects, as described by the EPPS21 parametrization \cite{Eskola:2021nhw}.  The corresponding predictions will be denoted NLO-BFKL + BLM hereafter. Moreover, for completeness of our analysis, we also will estimate the cross - sections considering the LO characteristic function, with the associated predictions being denoted by LO - BFKL.

\begin{figure}[t]
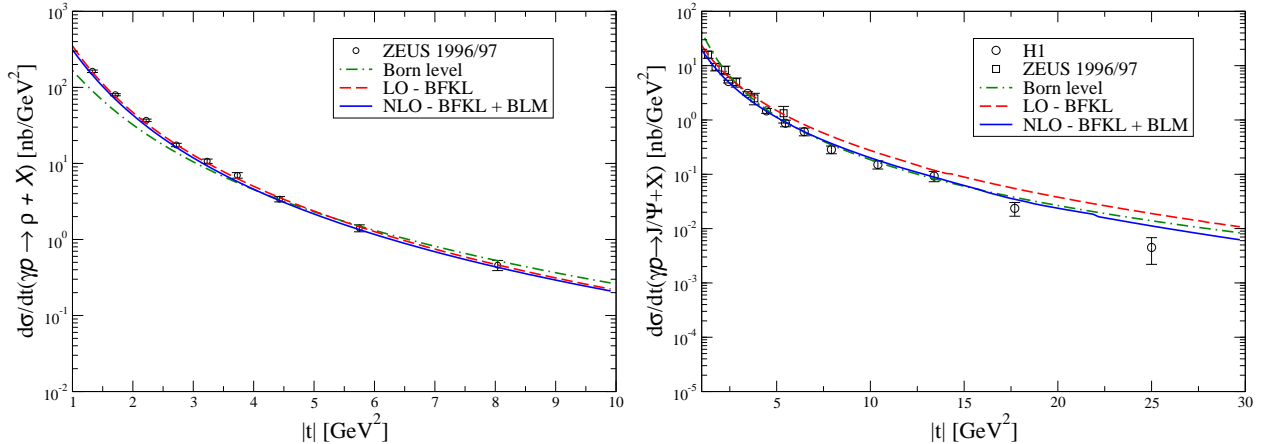

	\centering
	\begin{tabular}{ccccc}
\includegraphics[width=0.5\textwidth]{rho_dsdtnorm3.eps} &
\includegraphics[width=0.5\textwidth]{dsdtnorm_jpsi3.eps} 
	\end{tabular}
\caption{Comparison between the LO and NLO BFKL predictions and the HERA data  for the differential cross - sections associated with the $\rho$ (left panel) and $J/\Psi$ (right panel) photoproduction. Data from H1 \cite{H1:2003ksk}  and ZEUS \cite{ZEUS:2002vvv}  Collaborations. For completeness, the Born level predictions are also presented. }
\label{fig:distep}
\end{figure}

Initially, following Refs.  \cite{Forshaw:2001pf,Goncalves:2009gs,Goncalves:2010zb}, we will constrain the normalization of the LO and NLO predictions by adjusting the values of $\alpha_s$ in the impact factors, and  the constants  $\beta$ and $\delta$, present in the definition of the scale $\Lambda$, using the HERA data for the photoproduction of $\rho$ and $J/\Psi$ mesons at large - $t$ at a fixed photon - proton center - of - mass energy. For completeness, the results at the Born level, derived assuming $z = 0$ in Eq. (\ref{Eq:efe}), are also presented.  The comparison of our predictions with the HERA data \cite{H1:2003ksk,ZEUS:2002vvv} are presented in Fig. \ref{fig:distep}. A very good agreement with the  data is obtained using the values of $\alpha_s$, $\beta$ and $\delta$ presented in Table \ref{Tab:par}, which are
the values that we will use in our calculations for vector meson + jet production in UPHICs.


\begin{table}[t]
\begin{center}
\begin{tabular}{||l |r| r| r|| }
\hline
 {$\rho$}   & {\bf $\alpha_s$} & {\bf $\beta$ }  & {\bf $\delta$}\\
\hline
LO &  0.20 & 1.00 &  0.00 \\
 \hline
NLO & 0.29 & 0.00 & 0.10 \\
 \hline
 \hline
 $J/\Psi$ & & &  \\
\hline
LO &  0.20 & 0.65 &  0.25 \\ \hline
NLO & 0.20 & 0.00 & 0.10 \\
 \hline
\end{tabular}
\caption{Values of the parameters $\alpha_s$, $\beta$ and $\delta$, fixed by adjusting the predictions for the photoproduction of vector mesons at large - $t$  to the HERA data, and  used in  our calculations of the vector meson + jet photoproduction in UPHICs. }
\label{Tab:par}
\end{center}
\end{table}

\begin{figure}[t]
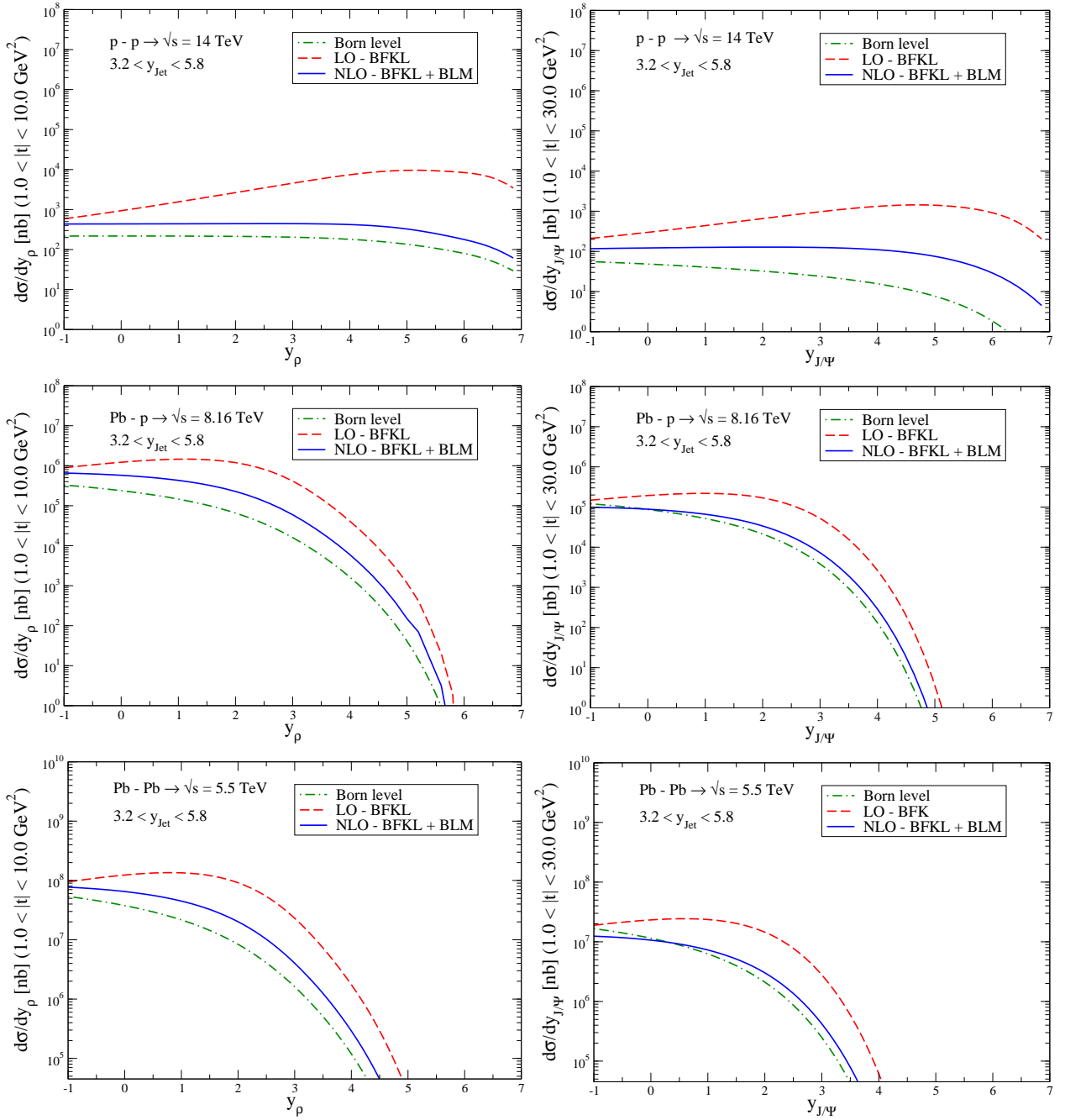

	\centering
	\begin{tabular}{ccccc}
\includegraphics[width=0.5\textwidth]{rho_dsdym_pp3.eps} &
\includegraphics[width=0.5\textwidth]{dsdym_pp3.eps} \\
\includegraphics[width=0.5\textwidth]{rho_dsdym_pPb3.eps} &
\includegraphics[width=0.5\textwidth]{jpsi_dsdym_pPb3.eps} \\
\includegraphics[width=0.5\textwidth]{rho_dsdym_PbPb3.eps} &
\includegraphics[width=0.5\textwidth]{dsdym_PbPb3.eps}  
	\end{tabular}
\caption{Rapidity distributions for the associated vector meson + jet photoproduction in $pp$ (upper panels), $Pbp$ (middle panels) and $PbPb$ (lower panels) collisions at the LHC. Results for $\rho$ (left panels) and $J/\Psi$ (right panels) derived integrating over the jet rapidity  and squared transverse momentum of the vector meson.}
\label{fig:dist}
\end{figure}

In our analysis of the  vector meson + jet photoproduction in UPHICs, we will explore the possibility that in a near future the FOCAL detector  could be used to measure the jet in the rapidity range $3.2 < y < 5.8$, with the associated meson being measured by a central or forward detector.  Considering this perspective, we will integrate the  rapidity of the jet in the differential cross-section, Eq. (\ref{Eq:dif}),  over the FOCAL rapidity range and present our predictions as a function of the vector meson rapidity. Moreover, we will integrate over $|t| = p_T^2$ in the range $1.0 \le |t| \le |t|_{max}$, with  $|t|_{max} = 10.0$ (30.0) GeV$^2$ for $\rho$ ($J/\Psi$) production, but we have verified that the predictions are similar for other values of the upper limit. The results for $pp$, $Pbp$ and $PbPb$ collisions at the LHC are presented in Fig. \ref{fig:dist}. As expected, the predictions for the $\rho$ production are larger than for the $J/\Psi$ one. Moreover, one has that $d\sigma (PbPb)  >  d\sigma (Pbp) > d\sigma (pp)$ due to the enhancements associated with the photon flux ($\propto Z^2$) and nuclear parton distributions ($\propto A$). In addition, the distribution is wider for $pp$ collisions due to the presence of more energetic photons in the flux associated with the proton in comparison to the nucleus ($\omega_{max} \propto 1/R$, where $R$ is the radius of the particle that emits the photon). Comparing the BFKL results with the Born level prediction, one has that the BFKL evolution implies the enhancement of the predictions with the increasing of the vector meson rapidities. However, such an increasing depends on the colliding system considered and the order that the BFKL kernel is estimated. As higher photon - hadron center - of - mass energies are probed in $pp$ collisions, the difference between the Born and BFKL results is larger in comparison to that predicted for $Pbp$ and $PbPb$ collisions. As the inclusion of the NLO corrections imply a modification on the energy dependence of the partonic differential cross-section $d\sigma(\gamma f \rightarrow Vf)/dt$, reducing its increasing with the energy, these corrections modify the corresponding rapidity distributions, as observed in Fig. \ref{fig:dist}. In comparison to the LO BFKL predictions,  the NLO corrections imply a decreasing of the normalization of the distributions, which become closer to the Born level predictions. 
Such results indicate that the rapidity distribution is sensitive to the description of the BFKL kernel.

\begin{figure}[t]
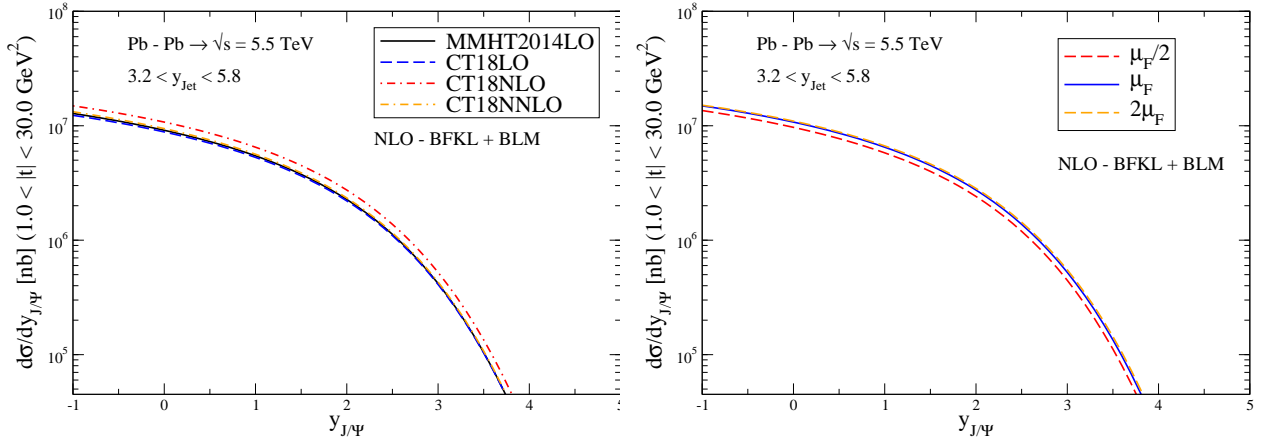

	\centering
	\begin{tabular}{ccccc}
\includegraphics[width=0.5\textwidth]{dsdym_PbPb3_pdfs.eps} &
\includegraphics[width=0.5\textwidth]{dsdym_PbPb3_scales.eps}
	\end{tabular}
\caption{Dependence on the PDFs (left panel) and factorization scale (right panel) assumed in the calculations of the rapidity distribution for the associated $J/\Psi$ + jet photoproduction in  $PbPb$ collisions at the LHC. }
\label{fig:pdfs_scales}
\end{figure}

{ A comment about the approach used in our analysis is in order. In our calculations, we have used the BLM prescription for incorporating (partial) next - to - leading corrections to the BFKL kernel. The main goal of this prescription is to reduce the instabilities in the resummed series and avoid that observables reach unphysical values when the rapidity separation between the produced particles becomes very large. However, the impact of this prescription is dependent on the observable considered. In recent years, the studies performed in Refs. \cite{Celiberto:2016hae,Celiberto:2017ptm,Bolognino:2018oth,Celiberto:2020wpk}   have demonstrated that the predictions for the total cross - sections associated with the light di-hadron and hadron - jet production have a substantial loss of statistical significance, due to the fact that the BLM prescription implies a value for the optimal renormalization scale that is significantly higher than the natural scales of the processes. Such results have motivated the proposition of other phenomenological approaches, where the high - energy resummed series is naturally stabilized by considering a combination of scales dictated by the kinematics of the process. Promising results have been obtained, e.g. in Refs. \cite{Celiberto:2020tmb,Celiberto:2021dzy,Celiberto:2022rfj,Celiberto:2022zdg}, which we refer the reader for a more detailed discussion. Such an alternative to stabilize the resummed series in the case of the process considered in this paper is a topic that we intend to explore in a forthcoming study. Here we will restrict our analysis to the BLM prescription. In order to quantify the impact of missing higher - order uncertainties in our predictions, in Fig. \ref{fig:pdfs_scales} we present the dependence of our results for the associated $J/\Psi$ + jet photoproduction in  $PbPb$ collisions at the LHC on the parameterization for the parton distribution functions (left panel) and factorization scale (right panel) assumed in the calculations. It is important to emphasize that we are only considering the central parametrizations provide by these groups, and that the uncertainty is larger if the predictions associated with the replicas are also shown. In addition, we are assuming that the factorization and renormalization scales are equal. One has that the predictions derived using different PDFs are similar (see left panel), which is expected since the typical values of $x_2$ probed at forward rapidities is large ($x_2 \gtrsim 0.1$), where the uncertainty in the PDFs is small. The larger effect is associated with the perturbative order  considered in the global analysis of the experimental data. In addition, the results presented in the right panel indicate that the predictions are slightly modified when the factorization scale is varied by a factor 2 around the central value.}


\section{Summary}

In this paper, we have investigated the BFKL effects on the vector meson + jet production in photon - induced interactions at the LHC. In particular, we have estimated the cross - sections considering the NLO corrections to the BFKL kernel and have explored the possibility that in a near future the FOCAL detector  could be used to measure the  jet at forward rapidities, with the associated meson being measured by a central or forward detector. We have presented predictions considering $pp$, $Pbp$ and $PbPb$ collisions and demonstrated that the associated cross-sections are non - negligible. Finally, our results indicated that 
a future experimental analysis of this process can be useful to constrain the description of the BFKL kernel and improve our understanding of the QCD dynamics.

\begin{acknowledgments}
{ We would like to thank the anonymous referee for many
useful suggestions and comments which have helped us
improve the manuscript.}
This work was partially supported by CAPES (Finance Code 001), CNPq, FAPERGS and INCT-FNA (Process No. 464898/2014-5). 

\end{acknowledgments}

\hspace{1.0cm}

\end{document}